\documentclass{svjour3}                     % onecolumn (standard format)
\usepackage{graphicx}

\usepackage{latexsym}
\usepackage{amssymb}

\journalname{NRL}
\begin{document}

\title{Graphene Bilayer Structures with Superfluid Magnetoexcitons}%

\author{A. A. Pikalov         \and
        D. V. Fil %etc.
}

\institute{A. A. Pikalov
           \and
           D. V. Fil \at
             Institute for Single Crystals, National Academy of Sciences of Ukraine, Lenin ave. 60 Kharkov 61001,
    Ukraine \\
              Tel.: +38-057-3410492\\
              \email{fil@isc.kharkov.ua}           %  \\
}

\date{Received: date / Accepted: date}

\maketitle

\begin{abstract}
We study superfluid behavior of a gas of spatially indirect
magnetoexcitons with reference to a system of two graphene layers
embedded in a multilayer dielectric structure. The system is
considered as an alternative of a double quantum well  in a GaAs
heterostructure. We determine a range of parameters (interlayer
distance, dielectric constant, magnetic field and gate voltage)
where magnetoexciton superfluidity can be achieved. Temperature of
superfluid transition is computed. A reduction of critical
parameters caused by impurities is evaluated and critical impurity
concentration is determined. \keywords{Graphene \and Exciton
superfluidity \and Multilayer heterostructures}
\end{abstract}

\section{Introduction}
\label{intro} Recent progress in creation of heterostuctures with
two graphene layers separated by a thin dielectrics  \cite{1a} opens
possibilities to use graphene for creation of multiple quantum well
structures with separately accessed  conducting layers. In \cite{1a}
SiO$_2$ substrate and  Al$_2$O$_3$ internal dielectric layer were
used. Another promising dielectric is hexagonal BN \cite{3a}. It has
a number of advantages, such as an atomically smooth surface that is
free of dangling bonds and charge traps, a lattice constant similar
to that of graphite, and a large electronic bandgap.

The attention to graphene heterostructures is caused,  in some part,
by the  idea to use them for a realization of superfluidity of
spatially indirect excitons \cite{g1,g2,g3,g4,g4a,g5,g6}. Bound
electron-hole pairs cannot carry electrical charge, but in bilayers
they can provide a flow of oppositely directed electrical currents.
Therefore, exciton superfluidity in bilayers should manifest itself
as a special kind of superconductivity - the counterflow one, that
means infinite conductance under a flow of equal in modulus and
oppositely directed currents in the layers.

The idea on counterflow superconductivity with reference to
electron-hole bilayers was put forward in \cite{1,2}. The attempts
to observe counterflow conductivity directly were done
\cite{e1,e2,e3} for bilayer quantum Hall systems   realized in GaAs
heterostructures. In the latter systems superconducting behavior
might be accounted for  magnetoexcitons \cite{moon,e-m}. The effect
is expected for the filling factors of Landau levels
$\nu_i=2\pi\ell^2 n_i$ ($\ell=\sqrt{\hbar c/eB}$ is magnetic length,
$n_i$ is the electron density in the $i$-th layer) satisfying the
condition $\nu_1+\nu_2=1$. The role of holes is played by empty
states in zero Landau level. In experiments \cite{e1,e2,e3} an
exponential increase of the counterflow conductivity under lowering
of temperature was observed, but zero-resistance state was not
achieved. The latter can be explained by the presence of unbound
vortices \cite{v1,v2,v3}. Such vortices may appear due to spatial
variation of the electron density
 caused by disorder.

To demonstrate counterflow superconductivity quantum Hall bilayers
 should have the parameters that satisfy two additional
conditions: $d \lesssim \ell$ and $\ell\lesssim a_B^*$, where $d$ is
the interlayer distance, and $a_B^*=\varepsilon \hbar^2/e^2 m^*$ is
the effective Bohr radius ($\varepsilon$ is the dielectric constant
of the matrix, and $m^*$ is the effective electron mass). The first
inequality comes from the dynamical stability condition. For
balanced bilayers ($\nu_1=\nu_2$) the mean-fields theory yields
$d<1.175 \ell$. The second inequality is the condition for the
Coulomb energy $e^2/\varepsilon\ell$ be smaller than the energy
distance between Landau levels. In GaAs $a_B^*\approx 10$ nm and the
condition $\ell\lesssim a_B^*$ is fulfilled at rather strong
magnetic fields $B\gtrsim 6$ T (actually, the experiments
\cite{e1,e2,e3} were done at smaller fields). At $d\lesssim 10$ nm
the interlayer tunneling is not negligible small and may result in a
locking of the bilayer for the counterflow transport at small input
current \cite{11,12}. At larger input current the system unlocks,
but the state becomes nonstationary one \cite{13,14,15} that is
accompanied by a dissipation (the power of losses is proportional to
the square of the amplitude of the interlayer tunneling
\cite{13,15}).

The idea to use graphene for the realization of electron-hole
superfluidity in quantum Hall bilayers \cite{g4,g4a,g5,g6} looks
very attractive. The distance between Landau Levels in monolayer
graphene is proportional to the inverse magnetic length, magnetic
field does not enter into the condition of smallness of the Coulomb
energy, and small magnetic fields can be used. Smaller magnetic
fields correspond to smaller critical temperature, but, at the same
time, they correspond to larger critical $d$. Use of large $d$
allows to suppress completely negative effects caused by interlayer
tunneling.

In this paper we concentrated on three questions. First, we
determine, in what range of internal parameters and external fields
magnetoexciton superfluidity can be realized. Second, we evaluate
critical temperature for pure system.  Third, we consider its
reduction caused by electron-impurity interaction. Our study extends
the results of \cite{g5}, where a system of two graphene layers
embedded into a bulk dielectric matrix was considered. Here we
investigate structures with one and two graphene layers situated at
the surface.

\section{Conditions for the electron-hole pairing in zero Landau level}
\label{sec1} Quantum Hall effect in graphene is characterized by
unusual systematics of Landau levels and the additional four-fold
degeneracy connected with two valleys and  two spin projections
\cite{rmp}. The energies of Landau levels in graphene are $E_{\pm
N}=\frac{\hbar v_F}{\ell}\sqrt{2|N|}$, where $N=0, 1, 2, \ldots$,
and $v_F\approx 10^6$ m/s is the Fermi velocity. In a free standing
graphene the $N=0$ Landau level is half-filled. A state with only
completely filled Landau levels corresponds to  a plateau at the
Hall conductivity plot (dependence of $\sigma_{xy}$ on electron
density). A free standing graphene is just between two plateaus
\cite{gr1}. A given quantum states in zero Landau is characterized
by the guiding center index $X$ and the combination of the spin and
valley indexes. Below we call four possible combinations, the
components, and numerate them by the index $\beta=1,2,3,4$.

We  describe electron-hole pairing in zero Landau level in graphene
by the wave function that is a generalization of the wave function
\cite{moon} to the multicomponent case
\begin{equation}\label{1}
|\Psi\rangle=\prod_\beta\prod_{X}(u_\beta c^+_{1\beta X}+v_\beta
c^+_{2\beta X})|0\rangle.
\end{equation}
Here $c^+_{i\beta X}$ is the electron creation operator (the
operator that fills a given state in $N=0$ Landau Level),
$|0\rangle$ is the state with empty zero level, $i$ is the layer
index. The $u-v$ coefficients satisfy the condition
$|u_\beta|^2+|v_\beta|^2=0$. The function (\ref{1}) can be rewritten
in the form
\begin{equation}\label{1-1}
|\Psi\rangle=\prod_\beta\prod_{X}(u_\beta +v_\beta c^+_{2\beta
X}h^+_{1\beta X})|vac1\rangle,
\end{equation}
where $h^+_{1\beta X}=c_{1\beta X}$ is the hole creation operator,
and the vacuum state is defined as
$|vac1\rangle=\prod_\beta\prod_{X}c^+_{1\beta X}|0\rangle$. One can
see that the function (\ref{1-1}) is an analog of the BCS function
in the Bardin-Cooper-Schrieffer  theory of superconductivity.

The quantity $\tilde{\nu}_\beta=|u_\beta|^2-|v_\beta|^2$ gives the
filling factor imbalance for the component $\beta$.  The order
parameter of the electron-hole pairing reads as
$\Delta_\beta=u^*_\beta v_\beta=\sqrt{1-\tilde{\nu}_\beta^2}e^{i
\varphi}/2$. If a given component is maximally imbalanced
($\tilde{\nu}_\beta=\pm 1$) the order parameter $\Delta_\beta$ is
equal to zero.

If a one component bilayer system is balanced, the order parameter
for the electron-hole pairing is maximum. But if the number of
components is even, the balance $\sum_\beta \tilde{\nu}_\beta=0$ can
be reached at $\tilde{\nu}_\beta=1$ for half of the components and
$\tilde{\nu}_\beta=-1$ for the other half. In the latter case all
$\Delta_\beta=0$. As is shown below, just such a state corresponds
to the energy minimum. In other words, in balanced graphene bilayers
electron-hole pairing does not occur.

\begin{figure}
  \includegraphics[width=7.5cm]{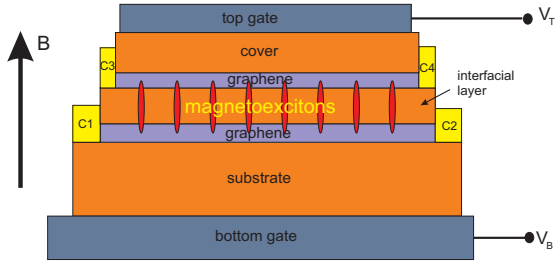}\\
  \caption{Schematic view of the system under study.
   C1 - C4 are the contacts. }\label{f1}
\end{figure}

At nonzero imbalance $\sum_\beta \tilde{\nu}_\beta\ne 0,\pm 2, \pm
4$   at least for one component $\tilde{\nu}_\beta\ne \pm 1$, and
electron-hole pairing may occur. Nonzero imbalance can be provided
by electrical field directed perpendicular to the layers. Such a
field can be created by a voltage difference applied between top and
bottom gates (see fig. \ref{f1}).

We consider the general structure "dielectric 1 - graphene 1 -
dielectric 2 - graphene 2 - dielectric 3"  with three different
dielectric constants $\varepsilon_1$, $\varepsilon_2$ and
$\varepsilon_3$. Dielectrics 1 and 3 are assumed to be thick (much
thicker than the distance between graphene layers $d$). Solving the
standard electrostatic problem we obtain the Fourier components of
the Coulomb interaction $V_{ii'}$ for the electrons located in $i$
and $i'$ graphene layers

\begin{equation}\label{2}
    V_{11}({\bf q})=\frac{4 \pi e^2}{q}
    \frac{\varepsilon_2+\varepsilon_3+(\varepsilon_2-\varepsilon_3)e^{-2 q
    d}}{(\varepsilon_2+\varepsilon_3)(\varepsilon_2+\varepsilon_1)-{(\varepsilon_2-\varepsilon_3)(\varepsilon_2-\varepsilon_1)}
    e^{-2 q d}},
\end{equation}
\begin{equation}\label{2-1}
    V_{22}({\bf q})=\frac{4 \pi e^2}{q}
    \frac{\varepsilon_2+\varepsilon_1+(\varepsilon_2-\varepsilon_1)e^{-2 q
    d}}{(\varepsilon_2+\varepsilon_3)
    (\varepsilon_2+\varepsilon_1)-{(\varepsilon_2-\varepsilon_3)(\varepsilon_2-\varepsilon_1)}
    e^{-2 q d}},
\end{equation}

\begin{equation}\label{3}
    V_{12}({\bf q})=\frac{8 \pi e^2}{q}
    \frac{{\varepsilon_2}e^{- q
    d}}{{(\varepsilon_2+\varepsilon_3)(\varepsilon_2+\varepsilon_1)}-{(\varepsilon_2-\varepsilon_3)(\varepsilon_2-\varepsilon_1)}
    e^{-2 q d}}.
\end{equation}

For electrons in $N=0$ Landau level in graphene the Hamiltonian of
Coulomb interaction has the form
\begin{eqnarray}\label{4}
    H_C=\frac{1}{2S}\sum_{i,i'}\sum_{X,X'}\sum_{\beta,\beta'}\sum_{\bf
    q}V_{ii'}({\bf q})e^{-\frac{q^2\ell^2}{2}+iq_x(X'-X)}\cr
    c^+_{i\beta X+q_y\ell^2/2}c^+_{i'\beta' X'-q_y\ell^2/2}
    c_{i'\beta' X'+q_y\ell^2/2}c_{i\beta X-q_y\ell^2/2},
\end{eqnarray}
where $S$ is the area of the system.
 The interaction with
the gate field is described by the Hamiltonian
\begin{equation}\label{4-1}
   H_G=-\frac{e V_g}{2}\sum_{X\beta}\left(c^+_{1\beta X} c_{1\beta
X}-c^+_{2\beta X} c_{2\beta X}\right),
\end{equation}
where $V_g$ is the interlayer voltage created by the external gate
(bare voltage).

Rewriting the wave function (\ref{1}) in the form
\begin{equation}\label{5}
|\Psi\rangle=\prod_X\prod_\beta\left(
\cos\frac{\theta_\beta}{2}c^+_{1\beta X}
+e^{i\varphi_\beta}\sin\frac{\theta_\beta}{2}c^+_{2\beta
X}\right)|0\rangle,
\end{equation}
and computing the  energy in the state (\ref{5}) we obtain
\begin{eqnarray}\label{6}
    E_{mf}=\frac{S}{8\pi\ell^2}\Bigg({W}\sum_{\beta\beta'}\cos \theta_\beta \cos
    \theta_{\beta'}- J_0\sum_\beta \cos^2 \theta_\beta
     -(2 e V_{g}+J_z)\sum_\beta \cos \theta_\beta\Bigg),
    \end{eqnarray}
where $W={e^2 d}/{\varepsilon_2 \ell^2}$ is the energy of direct
Coulomb interaction. The exchange interaction energies
$$J_{ik}=\frac{1}{2\pi}\int_0^{\infty}q V_{ik}(q) e^{-\frac{q^2 \ell^2}{2}}d q$$
determine the parameters $J_0=(J_{11}+J_{22})/2-J_{12}$ and
$J_z=J_{11}-J_{22}$. The relation between  $\theta_\beta$ and
$\tilde{\nu}_\beta$ is given by equation $\tilde{\nu}_\beta=\cos
\theta_\beta$.

Taking into account the inequalities $W>J_0$, and
$J_{11},J_{22}>J_{12}$ (that can be checked directly) we find that
at $V_g=0$ the minimum of (\ref{6}) is reached at
$\tilde{\nu}_1=\tilde{\nu}_2=1$, $\tilde{\nu}_3=\tilde{\nu}_4=-1$.
It indicates the absence of electron-hole pairing in balanced
systems.

If $V_g\ne 0$ and belongs to one of the intervals
\begin{equation}\label{ne1}
nW+J_{22}-J_{12}< eV_g< (n+2) W-J_{11}+J_{12},
\end{equation}
where $n=-4,-2,0,2$, the energy minimum is reached at
$\tilde{\nu}_{\beta_a}\ne \pm 1$ for one of the components. We will
call such a component the active one.

Let us, for instance, consider the interval (\ref{ne1}) with $n=0$.
Then the energy minimum is reached at
$$\tilde{\nu}_{\beta_a}=\frac{eV_g+J_z/2-W}{W-J_0}.$$
The case $\tilde{\nu}_{\beta_a}=0$ (with maximum order parameter)
corresponds to the voltage
\begin{equation}\label{6-2}
e V_g=-\frac{J_z}{2}+ W.
\end{equation}
Eq. (\ref{6-2}) determines the relation between magnetic field and
the gate voltage $V_g$. To keep $\tilde{\nu}_{\beta_a}=0$ the gate
voltage should be varied synchronically with $B$. In particular, at
$J_z=0$ ($\varepsilon_1=\varepsilon_3$) the quantities $V_g$ and $B$
are  linearly related:
\begin{equation}\label{7-1}
   V_g= \frac{\alpha d B c}{\varepsilon_2},
\end{equation}
 where $\alpha\approx 1/137$ is the fine structure
constant (the relation (\ref{7-1}) is given in SI units).

If only the gate voltage or magnetic field is varied, the order
parameter (and the critical temperature) changes nonmonotonically
reaching the maximum at the point determined by (\ref{6-2}).

\section{Collective mode spectrum and phase diagram}

The components that belong completely to one layer do not take part
in the pairing. In what follows we consider  the dynamics of only
the active component.

We describe the active component by the wave function
\begin{equation}\label{5-2}
|\Psi\rangle =\prod_X\left( \cos\frac{\theta_X}{2}c^+_{1, X+{Q_y
\ell^2}/{2}} +e^{i (Q_x
X+\tilde{\varphi}_X)}\sin\frac{\theta_X}{2}c^+_{2, X-{Q_y
\ell^2}/{2}}\right)|0\rangle
\end{equation}
(here and below we omit the component index). Eq. (\ref{5-2})
describes the state with nonzero counterflow currents.  To
illustrate this statement we neglect for a moment the order
parameter fluctuations ($\tilde{\varphi}_X=0$, $\theta_X=\theta_a$).

The order parameter is determined by the equation
\begin{eqnarray}\label{11}
    \Delta({\bf r})=\sum_{X,X'}\psi^*_X({\bf
    r})\psi_{X'}({\bf
    r})\langle\Psi|c^+_{1,X}c_{2,X'} |\Psi\rangle.
\end{eqnarray}
where $$\psi_X(\mathbf{r})=\frac{1}{\pi^{1/4}\sqrt{\ell L_y}}e^{-i
\frac{X y}{\ell^2}}e^{-\frac{(x-X)^2}{2\ell^2}}$$ is the
single-particle wave function in the coordinate representation,
$L_y$ is the width of the system. Substitution (\ref{5-2}) into
(\ref{11}) yields
\begin{equation}\label{11a}
 \Delta({\bf r})=\frac{\sin
\theta_a}{2}e^{-\frac{Q^2\ell^2}{2}}e^{i {\bf Q}\cdot{\bf r}}.
\end{equation}
One can see from Eq. (\ref{11a}) that ${\bf Q}=(Q_x, Q_y)$ is the
gradient of the phase of the order parameter.

Computing the energy in the state (\ref{5-2}) and neglecting the
fluctuations we obtain
\begin{equation}\label{e-mf}
    E_{0}=\frac{S}{8\pi \ell^2}\left(
    [W-F_S(0)]\cos^2 \theta_a
    - { F}_D(Q)\sin^2 \theta_a\right),
\end{equation}
where
\begin{eqnarray} \label{fs}
F_{S}(q)=\frac{1}{4\pi}\int_0^{\infty}p J_0(p q \ell^2)
[V_{11}(p)+V_{22}(p)]e^{-\frac{p^2 \ell^2}{2}}d p,
\end{eqnarray}
and
\begin{eqnarray} \label{fd}
F_{D}(q)=\frac{1}{2\pi}\int_0^{\infty}p J_0(p q \ell^2)
V_{12}(p)e^{-\frac{p^2 \ell^2}{2}}d p.
\end{eqnarray}

Electrical currents can be found from a variation of the energy
caused by a variation of the vector-potential
\begin{equation}\label{15}
    \delta E=-\frac{1}{c}\int d^2 r \sum_i{\bf j}_i \delta{\bf
    A}_i.
\end{equation}
Here ${\bf A}_i$ is the in-plane component of the vector-potential
in the layer $i$. To obtain the explicit expression for the
variation (\ref{15}) we replace the phase gradient in (\ref{e-mf})
with the gauge-invariant expression ${\bf Q}  -\frac{e}{\hbar
c}({\bf A}_{pl,1}-{\bf A}_{pl,2})$, where ${\bf A}_{pl,i}$ is the
parallel to the graphene layers component of the vector potential in
the layer $i$. Then, using (\ref{15}) one finds the currents
\begin{equation}\label{19}
    {\bf j}_1=-{\bf j}_2=-\frac{e}{\hbar}\frac{\sin^2 \theta_a}{8\pi\ell^2}\frac{d F_D(Q)}{d {\bf
    Q}}.
\end{equation}
At small gradients $Q\ell\ll 1$ Eq. (\ref{19}) is reduced to
\begin{equation}\label{19a}
    j_1=\frac{e}{\hbar}\rho_{s0}Q,
\end{equation}
where coefficient of proportionality between the current and the
phase gradient
\begin{equation}\label{20}
    \rho_{s0}=\frac{\ell^2}{32 \pi^2} \sin^2 \theta_a \int_0^{\infty}  p^3
    V_{12}(p) e^{-\frac{p^2 \ell^2}{2}} d p
\end{equation}
is called the zero temperature superfluid stiffness (the definition
is given in the next section). Since we neglect fluctuations, the
expression (\ref{19}) yields the current at $T=0$.

Implying the fluctuations of the amplitude and the phase of the
order parameter are small one can present the energy as
\begin{equation}\label{12}
    E=E_0+E_2+\ldots
\end{equation}
The quadratic in fluctuations term can be diagonalized:
\begin{eqnarray} \label{E2}
E_{2}=\sum_q [m_z(-q)K_{zz}(q)m_z(q)+
\frac{1}{4}\varphi(-q)K_{\varphi\varphi}(q)\varphi(q)
\cr-\frac{1}{2}(im_z(-q)K_{z\varphi}(q)\varphi(q)+c.c.)],
\end{eqnarray}
where
\begin{eqnarray} \label{mF} {m} _ {z} (q) = \frac{1}{2}\sqrt{\frac
{2 \pi l^2} {S}} \sum_X \left(\cos\theta_X-\cos \theta_a\right) e ^
{-{i} q X}, \cr {\varphi} (q) = \sqrt{\frac {2 \pi l^2} {S}} \sum_X
\tilde{\varphi} (X) e ^ {-{i} q X}
\end{eqnarray}
are the Fourier components of the fluctuations.

Eqs. (\ref{E2}) yields the energy of fluctuations with the wave
vector directed along the $x$ axis. The component of the matrix $K$
can be presented in form independent of the choice of the direction
of the coordinate axes
\begin{eqnarray} \label{Ks}
K_{zz}({\bf q},{\bf Q})=H({\bf q},{\bf Q})-F_S(|{\bf q}|)+F_D(|{\bf
Q}|)+\Xi({\bf q},{\bf Q})\cot^2\theta_a,\cr K_{\varphi\varphi}({\bf
q},{\bf Q})=\sin^2\theta_a \Xi({\bf q},{\bf Q}),\cr
K_{z\varphi}({\bf q},{\bf Q})=-\cos\theta_a [F_D(|\mathbf{q}
+\mathbf{Q}|)-F_D(|\mathbf{q} -\mathbf{Q}|)]/2,
\end{eqnarray}
where
\begin{eqnarray} \label{HFs}
H({\bf q},{\bf Q})=\frac{1}{2\pi\ell^2}\left[\frac{V_{11}({\bf
q})+V_{22}({\bf q})}{2} -V_{12}({\bf
q})\cos\left(|\mathbf{q}\times\mathbf{Q}|
\ell^2\right)\right]e^{-\frac{q^2 \ell^2}{2}}\cr \Xi({\bf q},{\bf
Q})=\left[F_D(|{\bf Q}|)- \frac{F_D(|\mathbf{q}
+\mathbf{Q}|)+F_D(|\mathbf{q} -\mathbf{Q}|)}{2}\right].
\end{eqnarray}
The quantities $K_{\alpha\beta}(q)$ in (\ref{E2}) are expressed in
terms of (\ref{Ks}) as $K_{\alpha\beta}(q)=K_{\alpha\beta}({\bf
q},{\bf Q})\Big|_{{\bf q}=q \mathbf{i}_x}$.

The quantity $\hbar \cos \theta_X/2$ can be treated as a
$z$-component of the pseudospin and it is canonically conjugated
with the phase $\varphi_X$. The Fourier transformed quantities
(\ref{mF}) are defined as canonical variables as well. The
 equations of motion for the quantities ${m} _ {z} (q)$ and
${\varphi} (q)$ read as
\begin{eqnarray} \label{13}
\hbar\frac{d {\varphi}(q)}{d t}= 2 K_{zz} (q) {m}_z(q)- i
K_{z\varphi}(q) {\varphi}(q) ,\cr \hbar\frac{d {m}_z(q)}{d t}=-
\frac{1}{2}K_{\varphi\varphi}(q) {\varphi}(q)  -  i
K_{z\varphi}(q) {m}_z(q).
\end{eqnarray}
Eqs. (\ref{13}) yield the collective mode spectrum
 $\Omega(q,{\bf
Q})=\sqrt{K_{\varphi\varphi}(q)K_{zz}(q)}+K_{z\varphi}(q)$.
Rotating the axes one obtains the excitation spectrum at general
${\bf q}$
\begin{equation}\label{e1}
    \Omega({\bf q},{\bf Q})=\sqrt{K_{\varphi\varphi}({\bf q},{\bf Q})K_{zz}({\bf q},{\bf Q})}
    +K_{z\varphi}({\bf q},{\bf Q}).
\end{equation}

At ${\bf Q}=0$ the spectrum (\ref{e1}) is isotropic. It can be
presented in the Bogolyubov form
\begin{equation}\label{e2}
    \Omega_0(q)=\sqrt{\varepsilon_q(\varepsilon_q+\gamma_q)}.
\end{equation}
In Eq. (\ref{e2})
\begin{equation}\label{eq0}
\varepsilon_q=F_D(0)-F_D(q)
\end{equation}
is the kinetic energy ($\varepsilon_q\approx\hbar^2 q^2/2M$ at
$q\ell\ll 1$, where $M$ is the magnetoexciton mass, see, for
instance \cite{mm}), and
\begin{equation}\label{gq}
\gamma_q=[H({\bf q},0)-F_S(q)+F_D(q)]\sin^2 \theta_0
\end{equation}
has the sense of the exciton-exciton interaction energy (that
includes the direct and exchange parts).

\begin{figure}
  \includegraphics[width=7.5cm]{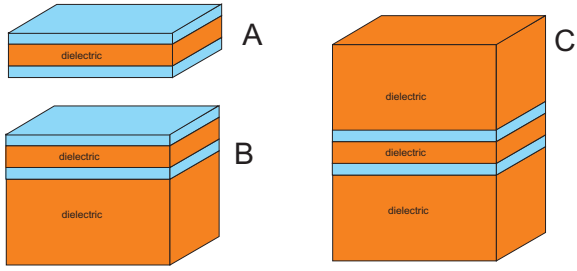}\\
  \caption{Graphene hererostructures under study.}\label{fig1}
\end{figure}

The condition for the dynamical stability of the state (\ref{5-2})
is the real valueness of the excitation spectrum (\ref{e2}). This
condition determines the diapason of  $d/\ell$ and $\varepsilon_i$
where  superfluid magnetoexciton state can be realized. To be more
concrete we consider three types of heterostructures. Type A is a
graphene-dielectric-graphene sandwich with two graphene layers at
the surface, type B is a graphene-dielectric-graphene-dielectric
structure with one such a layer, and type C is a system of two
graphene layers embedded in a dielectric matrix (Fig. \ref{fig1}).
For simplicity, we imply the same dielectric constants $\varepsilon$
for the interfacial layer and the substrate.

\begin{figure}
  \includegraphics[width=7.5cm]{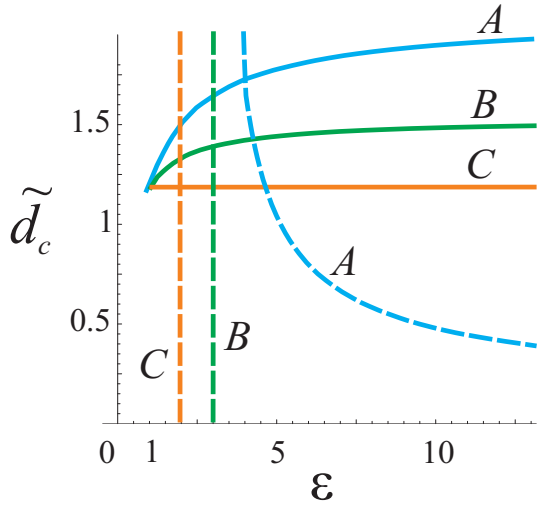}\\
  \caption{Phase diagram at $\tilde{\nu}_{a}=0$ for the graphene bilayers of A,B, and C type.
  Solid curves,
  $\widetilde{d}_c(\varepsilon)$;
  dashed curves, $\varepsilon_c(d/\ell)$.}\label{f2}
\end{figure}

The dynamical stability condition is fulfilled at $0<
d/\ell<\tilde{d}_c(\varepsilon)$, where $\tilde{d}_c(\varepsilon)$
depends on the imbalance parameter
$\tilde{\nu}_{\beta_a}\equiv\tilde{\nu}_{a}$. The dependence
$\tilde{d}_c(\varepsilon)$ at $\tilde{\nu}_{a}=0$ is shown in Fig.
\ref{f2}.

The requirement for the Coulomb energy be smaller than the distance
between Landau levels yields the restriction on $\varepsilon$. Since
we study the pairing in $N=0$ Landau level we  compare the Coulomb
energy with the energy distance between $N=0$ and $N=1$ levels
$\omega_c=\sqrt{2} \hbar v_F/\ell$.

 We have
four parameters that characterize the Coulomb energy $W$, $J_{11}$,
$J_{22}$ and $J_{12}$. At $d/\ell<\tilde{d}_c$ the largest of them
is $J_{11}$ (the intralayer exchange interaction in the graphene
layer at the surface). Therefore, it is natural to consider the
condition
\begin{equation}\label{14}
    J_{11}<\omega_c
\end{equation}
as the additional restriction on the parameters. Eq. (\ref{14}) can
be rewritten as $\varepsilon>\varepsilon_c(d/\ell)$. The quantity
$\varepsilon_c$ can be understood as a critical dielectric constant.
The dependence $\varepsilon_c(d/\ell)$ is also shown in Fig.
\ref{f2}.

Two conditions $d/\ell<\tilde{d}_c(\varepsilon)$ and
$\varepsilon>\varepsilon_c(d/\ell)$ determine the range of
parameters where one can expect a realization of electron-hole
pairing and magnetoexciton superfluiduty in graphene bilayer
systems.

\section{Critical  temperature }

In a bilayer graphene heterostructure with a fixed $d$ the
magnetoexciton superfluidity can be realized in a wide range of
magnetic field. Variation of $B$ at fixed gate voltage results in a
change of imbalance of the active component. Simultaneous tuning of
$V_g$ allows to keep zero imbalance $\tilde{\nu}_{a}=0$ and maximum
order parameter under variation of $B$. In this section we study the
dependence of
 critical temperature on magnetic field
implying such a simultaneous tuning.

Superfluid transition temperature is given by the
Berezinskii-Kostelitz-Thouless equation \cite{moon}
\begin{equation}\label{15-1}
    T_c=\frac{\pi}{2}\rho_s(T_c),
\end{equation}
where $\rho_s(T)$ is the superfluid stiffness at finite temperature.
The superfluid stiffness is defined as the coefficient in the
expansion of the free energy in the phase gradient $F=F_0+\int d^2 r
\rho_s (\nabla \varphi)^2/2$. In a weakly nonideal Bose gas it is
equal to $\rho_s=\hbar^2 n_s/m$, where $n_s$ is the superfluid
density. As was shown in  previous section, superfluid stiffness
determines also the supercurrent.

  Taking into account linear excitations we present
 the free energy $F=E_{0}-TS$ in the following form
 \begin{equation}\label{21}
F=E_0+T\sum_{\bf q}\ln\left(1-e^{-\frac{\Omega({\bf q},{\bf
Q})}{T}}\right).
\end{equation}
Expansion of Eq. (\ref{21}) yields the following expression for the
superfluid stiffness
\begin{equation}\label{22}
    \rho_{s}(T)=\rho_{s0}+\frac{1}{S}\sum_{\bf q}\left[\left(\frac{d^2 \Omega({\bf q},{\bf Q})}
    {d Q^2}\right)\Bigg|_{Q=0}N_q-\frac{1}{T}N_q(1+N_q)
    \left(\frac{d \Omega({\bf q},{\bf Q})}{d
    Q}\right)^2\Bigg|_{Q=0}\right].
\end{equation}
It follows from (\ref{22}) and (\ref{e1}) that
$\rho_{s}(T)<\rho_{s0}$  (thermal fluctuations reduce the superfluid
stiffness).

For the spectrum $\Omega({\bf q})=E(q)+\hbar {\bf q}{\bf v}$ (where
${\bf v}=\hbar\nabla \varphi/m$ is the superfluid velocity)
(\ref{22}) yields the well-known answer for the superfluid density
\cite{lp}. Eq. (\ref{22}) generalizes the results \cite{lp} for the
general case.

\begin{figure}
  \includegraphics[width=7.5cm]{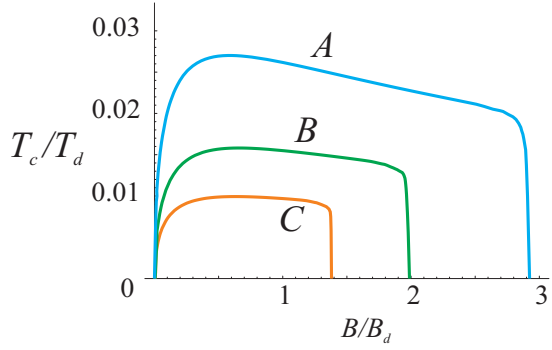}\\
  \caption{Critical temperature vs
magnetic field for  A, B,  and C structures. Temperature is given in
units of $T_d=e^2/\varepsilon d$, magnetic field, in units of
$B_d=\phi/\pi d^2$.}\label{f4}
\end{figure}

The dependence of critical temperature on  magnetic field  at
$\tilde{\nu}_{a}=0$ and $\varepsilon=4$ is shown in Fig. \ref{f4}.
One can see that the maximum critical temperature is reached
approximately at $B\approx 0.5 B_d$, where $B_d=\phi/\pi d^2$ with
$\phi=hc/2e$, the magnetic flux quantum.

\section{Influence of impurities on the  critical parameters}
\label{s4}

In the previous section we have determined the influence of thermal
fluctuations on the superfluid stiffness. In this section we
consider the effect of reduction of the superfluid stiffness caused
by the interaction of magnetoexcitons with impurities.

The Hamiltonian of the interaction of the active component with
impurities can be presented in the form
\begin{equation}\label{23}
   H_{imp}=\frac{1}{2 S}\sum_{\bf q}  U_{z}({\bf q})
   \left(\hat{\rho}_1({\bf q})- \hat{\rho}_2({\bf q})\right),
\end{equation}
where $U_z({\bf q})=U_1({\bf q})-U_2({\bf q})$, $U_i({\bf q})$ is
the Fourier-component of the impurity potential in the layer $i$,
and
\begin{equation}\label{24}
    \hat{\rho}_i({\bf q})=\sum_X c^+_{i,X+\frac{q_y \ell^2}{2}}
    c_{i,X-\frac{q_y \ell^2}{2}}\exp\left(-i q_x X-\frac{q^2
    \ell^2}{4}\right)
\end{equation}
is the Fourier component of the electron density operator for the
active component.

In the state (\ref{5-2}) the energy of interaction with the
impurities expressed in terms of $m_z(q)$ reads as
\begin{equation}\label{26}
    E_{imp}=const+\sum_{q}
    \tilde{U}_z(q)m_z(q),
    \end{equation}
where $$\tilde{U}_z(q)=\frac{1}{\sqrt{2\pi\ell^2S}}U_z(q {\bf
i}_x)e^{-\frac{q^2\ell^2}{4}}.$$

The interaction (\ref{26}) induces the fluctuations of the density
and the phase of the order parameter. Their values can be obtained
from the Euler-Lagrange equations
\begin{eqnarray}\label{EL}
  \frac{\delta E}{\delta m_z(q)}=0,\cr
  \frac{\delta E}{\delta \varphi(q)}=0,
\end{eqnarray}
where $E$ is the energy of the system, described by the
Hamiltonian $H=H_C+H_G+H_{imp}$ in the state (\ref{5-2}).

Eqs. (\ref{EL}) solved in linear in impurity potential approximation
yield
\begin{eqnarray}\label{12-a}
m_z(q)=-\frac{1}{2}\frac{K_{\varphi\varphi}(q) \tilde{U}_{z}(q {\bf
\hat{x}})} {K_{zz}(q)
K_{\varphi\varphi}(q)-K_{z\varphi}^2(q)},\\\label{12-b}
\varphi(q)=\frac{i K_{z\varphi}(q) \tilde{U}_{z}(q{\bf \hat{x}})
}{K_{zz}(q) K_{\varphi\varphi}(q)-K_{z\varphi}^2(q)}.
\end{eqnarray}
Substituting (\ref{12-a}), (\ref{12-b}) into the expression for the
energy one finds the correction to the energy caused by the
electron-impurity interaction
\begin{equation}\label{dE}
    \Delta E= -\frac{1}{4}\sum_q \tilde{U}_z(q)\tilde{U}_z(-q)\frac{
K_{\varphi\varphi}(q)}{K_{zz}(q)
K_{\varphi\varphi}(q)-K_{z\varphi}^2(q)}.
\end{equation}
In Eq. (\ref{dE}) the contribution of fluctuations with the wave
vectors directed along $x$ is taken into account. Summing the
contribution for all wave vectors one obtains
\begin{equation}\label{dE1}
    \Delta E= -\frac{1}{8\pi\ell^2 S}\sum_{\bf q} U_z({\bf q}) U_z(-{\bf q})\frac{
K_{\varphi\varphi}({\bf q}, {\bf
Q})e^{-\frac{q^2\ell^2}{2}}}{K_{zz}({\bf q}, {\bf Q})
K_{\varphi\varphi}({\bf q}, {\bf Q})-K_{z\varphi}^2({\bf q}, {\bf
Q})}.
\end{equation}

For simplicity, we specify the case where impurities are located
in graphene layers. Then the Fourier-component of the impurity
potential can be presented in the form
\begin{equation}\label{ip}
   {U}_{z}({\bf q})=\sum_{a}  e^{i {\bf q r}_{a} } u_{z,i}({\bf q}),
\end{equation}
where ${\bf r}_{a}$ are the impurity coordinates, and $u_{z,i}({\bf
q})=u_{1,i}({\bf q})-u_{2,i}({\bf q})$ with $u_{k,i}({\bf q})$, the
potential in the layer $k$ of a single impurity centered at ${\bf
r}=0$ in the layer $i$.

Averaging over impurities yields
\begin{equation}\label{ip1}
    \Delta E=-\frac{n_{imp}}{8\pi\ell^2}\sum_{\bf q} \left(|u_{z,1}({\bf q})|^2+|u_{z,2}({\bf q})|^2\right)
     \frac{
K_{\varphi\varphi}({\bf q}, {\bf
Q})e^{-\frac{q^2\ell^2}{2}}}{K_{zz}({\bf q}, {\bf Q})
K_{\varphi\varphi}({\bf q}, {\bf Q})-K_{z\varphi}^2({\bf q}, {\bf
Q})},
\end{equation}
where $n_{imp}$ is the impurity concentration in a layer.

At $Q\ell\ll 1$ the energy (\ref{ip1}) can be  expanded in series as
\begin{equation}\label{ip2}
    \Delta E=\Delta E_0+\frac{S}{2}\Delta\rho_{s}Q^2,
\end{equation}
where
\begin{equation}\label{drs}
    \Delta\rho_s=\frac{n_{imp}}{8\pi\ell^2 S}\sum_{\bf q}
    \frac{\left(|u_{z,1}({\bf q})|^2+|u_{z,2}({\bf q})|^2\right)e^{-\frac{q^2\ell^2}{2}}}
{{K_{zz}^2({\bf q}, 0)}}\left(\frac{\partial^2
    K_{zz}({\bf q}, {\bf Q})}{\partial Q^2}
    \Big|_{Q=0}-\frac{2\left(\frac{\partial
    K_{z\varphi}({\bf q}, {\bf Q})}{\partial Q}
    \Big|_{Q=0}\right)^2}{K_{\varphi\varphi}({\bf q}, 0)}\right)
\end{equation}
is the correction of the superfluid stiffness. One can check that
the correction $\Delta\rho_s$ is negative. Thus, the interaction
with impurities results in decrease of critical parameters.

At $\tilde{\nu}_{a}=0$ Eq. (\ref{drs}) is reduced to
\begin{equation}\label{dt1}
 \Delta\rho_s=-\frac{ n_{imp}}{S}\sum_{\bf q}\frac{\left(|u_{z,1}({\bf q})|^2+|u_{z,2}({\bf q})|^2\right)
 e^{-\frac{q^2\ell^2}{2}}} {K_{zz}^2({\bf q}, 0)}
 \left[\rho_{s0}-\frac{q^2
V_{12}(q)e^{-\frac{q^2\ell^2}{2}}}{32\pi^2}\right].
\end{equation}
where $\rho_{s0}$ (Eq. (\ref{20})) is taken at $\theta_a=\pi/2$.

\begin{figure}
  \includegraphics[width=7.5 cm]{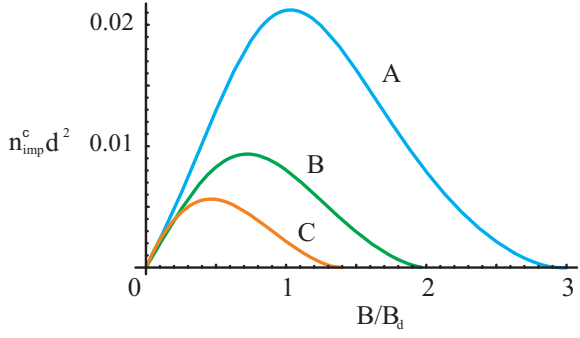}\\
  \caption{Critical impurity concentration vs magnetic field
  for charged impurities located in graphene layers.} \label{f6}
\end{figure}

 The shift of critical temperature is
evaluated as
 $\Delta T_c/T_c\approx \Delta
\rho_s/\rho_{s0}$\footnote{Since in our approach we assume smallness
of $\Delta \rho_s/\rho_{s0}$ it is just an estimate}. We define the
critical impurity concentration $n_{imp}^c$ as a concentration at
which $\Delta\rho_s/\rho_{s0}=1$ . We consider charged impurities
with the potential $u_{z,i}({\bf q})=(-1)^i(V_{12}({\bf
q})-V_{ii}({\bf q}))$. The dependence of critical impurity
concentration on magnetic field at $\varepsilon=4$  and
$\tilde{\nu}_{a}=0$ is shown in Fig. \ref{f6}. We also evaluated
critical concentrations for neutral impurities. These concentrations
are much larger, and the influence of neutral impurities can be
neglected.

\section{Conclusion}
\label{concl}

In conclusion, we present some  estimates. Let us specify the type B
structure (the one used in \cite{1a}) with $d=20$ nm and
$\varepsilon=4$. For this structure the maximum critical temperature
$T_c\approx 3$ K (in pure case) is reached in magnetic field
$B\approx 0.8$ T. At such $B$ the critical impurity concentration is
$n_{imp}^c \approx 2 \cdot 10^9$ cm$^{-2}$. The gate voltage
determined by Eq. (\ref{6-2}) is $V_g\approx 6$ mV, that corresponds
to electrostatic field $E\approx 3$ kV/cm.

Basing on the results of our study  we may state the following.

1. Graphene bilayer structures are perspective objects for the
observation of magnetoexciton superfluidity.
    The advantages are smaller magnetic fields and no restriction
    from above on physical interlayer distance, that means the possibility to suppress completely
    interlayer tunneling.

2. Gate voltage should be created between graphene layers for a
realization of magnetoexciton superfluidity.

3. Certain conditions on dielectric constant and on the ratio
between interlayer distance and magnetic length should be satisfied.

4. Structures with graphene layers situated at the surface have
larger critical parameters.

5. Neutral impurities are not dangerous for the magnetoexciton
superfluidity, but the concentration of charged impurities should be
controlled.

\begin{acknowledgements}
This study is supported by the Ukraine State Program
"Nanotechnologies and nanomaterials" Project No 1.1.5.21.
\end{acknowledgements}


\begin{thebibliography}{3}

\bibitem{1a} Kim S, Jo I, Nah J, Yao Z, Banerjee SK, Tutuc E:
{\bf Coulomb drag of massless fermions in graphene}. {\it Phys. Rev.
B}  2011, {\bf 83}: 161401(1-4).
\bibitem{3a} Dean CR, Young AF, Meric I, Lee C, Wang L, Sorgenfrei S, Watanabe K, Taniguchi T,
Kim P, Shepard KL, Hone J: {\bf Boron nitride substrates for
high-quality graphene electronics}. {\it Nature Nanotechnology} 2010
{\bf 5}: 722-726.


\bibitem{g1}
    Lozovik YE, Sokolik AA: {\bf Electron-hole pair condensation in a graphene
    bilayer}. {\it JETP Letters} 2008, {\bf 87}(1): 55-59.
\bibitem{g2} Min H, Bistritzer R, Su JJ and MacDonald AH: {\bf Room-temperature superfluidity in graphene bilayers}. {\it Phys.
Rev. B} 2008,  {\bf 78}: 121401(1-4).

\bibitem{g3}    Seradjeh B, Weber H, Franz M: {\bf Vortices,
Zero Modes, and Fractionalization in the Bilayer-Graphene Exciton
Condensate}. {\it Phys. Rev. Lett.} 2008, {\bf 101}(24):
246404(1-4).
\bibitem{g4} Berman OL, Lozovik YE, Gumbs G:
{\bf Bose-Einstein condensation and superfluidity of magnetoexcitons
in bilayer graphene}. {\it Phys. Rev.} B 2008, {\bf 77}:
155433(1-10).
\bibitem{g4a}
Lozovik YE, Merkulova SP, Sokolik AA: {\bf Collective electron
phenomena in graphene} {\it Phys. Usp.} 2008, {\bf 51}(7): 727-744.
\bibitem{g5} Fil DV, Kravchenko LY: {\bf
Superconductivity of electron-hole pair in a bilayer graphene system
in a quantizing magnetic field}. {\it Low Temp. Phys.} 2009, {\bf
35}: 712-723.

\bibitem{g6} Bezuglyi, AI: {\bf Dynamical
equation for an electron-hole pair condensate in a system of two
graphene layers}. {\it Low Temp. Phys.} 2010, {\bf 36}: 236-242.


\bibitem{1} Lozovik YE, Yudson VI: {\bf Novel mechanism of superconductivity--
pairing of spatially separated electrons and holes}. {\it Sov. Phys.
JETP} 1976, {\bf 44}: 389-397.
\bibitem{2} Shevchenko SI: {\bf Theory of superconductivity of systems
with pairing of spatially separated electrons and holes}. {\it Sov.
J. Low Temp. Phys.} 1976, {\bf 2}: 251-256.

\bibitem{e1} Kellogg M,
Eisenstein JP, Pfeiffer LN, West KW: {\bf Vanishing Hall Resistance
at High Magnetic Field in a Double-Layer Two-Dimensional Electron
System}. {\it Phys. Rev. Lett.} 2004, {\bf 93}: 036801(1-4).
\bibitem{e2} Wiersma RD, Lok JGS, Kraus S, W. Dietsche W, von
Klitzing K, Schuh D, Bichler M, Tranitz HP, Wegscheider W: {\bf
Activated Transport in the Separate Layers that Form the $\nu_T=1$
Exciton Condensate}. {\it Phys. Rev. Lett.} 2004, {\bf 93}:
266805(1-4).
\bibitem{e3} Tutuc E, Shayegan M, Huse DA:
{\bf Counterflow Measurements in Strongly Correlated GaAs Hole
Bilayers: Evidence for Electron-Hole Pairing}, {\it Phys. Rev.
Lett.} 2004, {\bf 93}: 036802(1-4).


\bibitem{moon} Moon K, Mori H, Yang K, Girvin SM, MacDonald AH,
Zheng L, Yoshioka D, and Zhang SC {\bf Spontaneous interlayer
coherence in double-layer quantum Hall systems: Charged vortices and
Kosterlitz-Thouless phase transitions}. {\it Phys. Rev.} B 1995,
{\bf 51}: 5138-5170.

\bibitem{e-m} Eisenstein JP, MacDonald AH:
{\bf Bose-Einstein condensation of excitons in bilayer electron
systems}. {\it Nature} 2004, {\bf 432}: 691-694.

\bibitem{v1}Huse DA: {\bf Resistance due to vortex motion in the $\nu=1$ bilayer quantum Hall superfluid}.
 {\it Phys. Rev. } B 2005, {\bf 72}:
064514(1-4).
\bibitem{v2}Roostaei B, Mullen KJ, Fertig HA, Simon SH: {\bf Theory of Activated Transport in Bilayer Quantum Hall Systems}. {\it Phys.
Rev. Lett.} 2008, {\bf 101}: 046804(1-4).
\bibitem{v3}Fil DV, Shevchenko SI:
{\bf Transport properties of $\nu=1$ quantum Hall bilayers. Phenomenological description}. {\it Phys. Lett. }A 2010, {\bf 374}:
3335-3340.

\bibitem{11} Yoon Y, Tiemann L, Schmult S, Dietsche W, von Klitzing K, Wegscheider W:
{\bf Interlayer Tunneling in
Counterflow Experiments on the Excitonic Condensate in Quantum Hall
Bilayers}. {\it Phys. Rev. Lett.} 2010, {\bf 104}: 116802(1-4).

\bibitem{12} Fil DV:
{\bf Locking and unlocking of the counterflow transport in $\nu=1$
quantum Hall bilayers by tilting of magnetic field}. {\it Phys.
Rev.} B 2010, {\bf 82}: 193303(1-4).
\bibitem{13} Fil DV, Shevchenko SI: {\bf Interlayer tunneling and the problem of
superfluidity in bilayer quantum Hall systems}. {\it Low Temp.
Phys.} 2007, {\bf 33}: 780-782.
\bibitem{14} Su JJ, MacDonald AH:
{\bf How to make a bilayer exciton condensate flow}. {\it Nature Physics} 2008, {\bf 4}:
799-802.
\bibitem{15}  Fil DV, Shevchenko SI:
{\bf Josephson vortex motion as a source for dissipation of
superflow of e-h pairs in bilayers}. {\it J. Phys.: Condens. Matter}
2009, {\bf 21}: 215701(1-9).
\bibitem{rmp} Castro Neto AH, Guinea F, Peres NM, Novoselov KS,
Geim AK: {\bf The electronic properties of graphene}. {\it Rev. Mod.
Phys.} 2009, {\bf 81}: 109-162.

\bibitem{gr1}Novoselov KS, McCann E, Morozov SV,  Fal'ko VI,  Katsnelson MI, Zeitler U,
Jiang D, Schedin F, Geim AK: {\bf Unconventional quantum Hall effect
and Berry's phase of $2\pi$ in bilayer graphene.} {\it Nature
Physics} 2006, {\bf 2}: 177 - 180.
\bibitem{mm} Kravchenko LY, Fil DV: {\bf Critical currents and giant non-dissipative
drag for superfluid electron–hole pairs in quantum Hall
multilayers.} {\it J. Phys.: Condens. Matter} 2008, {\bf 20}:
325325(1-9).

\bibitem{lp} Lifshitz EM, Pitaevskii LP: Statistical Physics, Part
2. Pergamon Press, 1980.


\end{thebibliography}
\end{document}